\documentclass[12pt]{article}
\usepackage{graphics,epsfig}
\usepackage[english]{babel}
\newcommand{\cinst}[2]{$^{\mathrm{#1}}$~#2\par}
\newcommand{\crefi}[1]{$^{\mathrm{#1}}$}

\begin{document}


\begingroup

\raisebox{0.5cm}[0cm][0cm] {
\begin{tabular*}{\hsize}{@{\hspace*{5mm}}ll@{\extracolsep{\fill}}r@{}}
\begin{minipage}[t]{3cm}
\vglue.5cm
\end{minipage}
&
\begin{minipage}[t]{7cm}
\vglue.5cm
\end{minipage}
&

\end{tabular*}
}

\begin{center}

{\Large{\bf A study of the nuclear medium influence\\
 on transverse momentum of hadrons\\
\vspace{0.22cm}

produced in deep inelastic neutrino scattering }}
\end{center}

\vspace{1.cm}

\begin{center}
{\large SKAT Collaboration}

 N.M.~Agababyan\crefi{1}, V.V.~Ammosov\crefi{2},
 M.~Atayan\crefi{3},\\
 N.~Grigoryan\crefi{3}, H.~Gulkanyan\crefi{3},
 A.A.~Ivanilov\crefi{2},\\ Zh.~Karamyan\crefi{3},
V.A.~Korotkov\crefi{2}

\setlength{\parskip}{0mm}
\small

\vspace{1.cm} \cinst{1}{Joint Institute for Nuclear Research,
Dubna, Russia} \cinst{2}{Institute for High Energy Physics,
Protvino, Russia} \cinst{3}{Yerevan Physics Institute, Armenia}
\end{center}
\vspace{100mm}

{\centerline{\bf YEREVAN  2003}}

\newpage
\begin{abstract}

\noindent The influence of nuclear effects on the transverse
momentum $(p_T)$ distributions of neutrinoproduced hadrons is
investigated using the data obtained with SKAT propane-freon 
bubble chamber irradiated in the neutrino beam 
(with $E_{\nu}$ = 3-30 GeV) at Serpukhov accelerator. 
Dependences of $<p_T^2>$ on the kinematical variables of 
inclusive deep-inelastic scattering and of
the produced hadrons are measured. It has been observed, that the
nuclear effects cause an enhancement of $<p_T^2>$ of hadrons (more
pronounced for the positively charged ones) produced in the target
fragmentation region at low invariant mass of the hadronic
system (2 $< W <$ 4 GeV) or at low energies transferred to the
current quark (2 $< \nu < 9$ GeV). At higher $W$ or $\nu$, no
influence of nuclear effects on $<p_T^2>$ is observed. 
Measurement results are compared with predictions of a
simple model, incorporating secondary intranuclear
interactions of hadrons (with a formation length extracted from
the Lund fragmentation model), which qualitatively reproduces the
main features of the data.
\end{abstract}


\newpage
\section{Introduction}

The experimental study of the hadron production in deep inelastic
scattering (DIS) of leptons on nuclei is an important source of
information on the space-time evolution of the leptoproduced quark
strings. Depending on the features of the latter, the nuclear
medium influences differently the inclusive spectra of final
hadrons, in particular, affecting their yields and the mean
transverse momentum in various domains of the phase space.
Hitherto  no detailed data inferred in the same experiment are
available for transverse momentum distributions in the
lepton-nucleus DIS. This gap is partly filled by the present work,
where the influence of the nuclear effects on the transverse
momentum of neutrinoproduced hadrons is explored using the data
collected in the neutrino experiment with bubble chamber SKAT. In Section 2,
the experimental procedure is briefly described. The experimental
results are presented in Section 3, discussed in Section 4 and
summarized in Section 5.

\section{Experimental procedure}

The experiment was performed with SKAT bubble
chamber \cite{ref1}, exposed to a wideband neutrino beam obtained
with a 70 GeV primary protons from the Serpukhov accelerator. The
chamber was filled with a propan-freon mixture containing 87 vol\%
propane ($C_3H_8$) and 13 vol\% freon ($CF_3Br$) with the
percentage of nuclei H:C:F:Br = 67.9:26.8:4.0:1.3 \%. A 20 kG
uniform magnetic field was provided within the operating chamber
volume.\\ Charged current interactions, containing a negative muon
with momentum $p_{\mu}>$ 0.5 GeV are selected. Other negatively
charged particles are considered to be $\pi^-$ mesons. Protons
with momentum below 0.6-0.65 GeV$/c$ and a fraction of protons
with momentum up to 0.85 GeV$/c$ were identified by their stopping
in the chamber. More details concerning the experimental
procedure, in particular, the event selection criteria and the
reconstruction of the neutrino energy $E_{\nu}$ can be found in
our previous publications \cite{ref2,ref3,ref4,ref5}. 
Each event is given a weight (depending on the charged
particle multiplicity) which corrects for the fraction of events
excluded due to improperly reconstruction. \\ For further analysis the
events with 3 $<E_{\nu}<$ 30 GeV were accepted, provided that the
invariant mass of the hadronic system $W>$ 2 GeV and the
transfer momentum squared $Q^2>$ 1 (GeV$/c)^2$. The full sample
consists of 2222 events (3167 weighted events). The mean values
of the kinematical variables are: $<E_{\nu}>$ = 10.8 GeV, $<Q^2>$ =
3.6 (GeV/$c)^2$, $<W>$ = 3.0 GeV, $<W^2>$ = 9.5 GeV$^2$, and, for
the energy ${\nu}$ transferred to the hadronic system, $<{\nu}>$ =
6.5 GeV. \\ The whole event sample is subdivided,
using several topological and kinematical criteria
\cite{ref4,ref5}, into three subsamples: the 'cascade' subsample
$B_S$ with a sign of intranuclear secondary interactions, the
'quasiproton' $(B_p)$ and 'quasineutron' $(B_n)$ subsamples, the
latters two composing the 'quasinucleon' subsample $(B_N \equiv B_p +
B_n)$ for which no sign of secondary interactions is observed.
The corresponding event numbers for $B_S, B_p$ and $B_n$
subsamples are 1190, 477 and 555 (weighted numbers $N_S$ =
1736, $N_p$ = 680 and $N_n$ = 751), respectively.
\\ About 40\% of subsample $B_p$ is contributed by interactions
with free hydrogen. Weighting the 'quasiproton' events with
a factor of 0.6, one can compose a 'pure' nuclear sample: $B_A = B_S
+ B_n + 0.6 B_p$ (with an effective atomic weight $\bar{A}$ = 28).
It could be also shown \cite{ref4,ref5}, that a subsample composed
as $B_D = B_n + 0.6 B_p$ approximately corresponds to the
neutrino-deuterium interactions.

\section{Experimental results}

In present analysis the identified protons are not included. All
rest positively (labelled as $h^+$) and negatively (labelled as
${\pi}^-$) charged hadrons are given the pion mass $m_{\pi}$. 
The transverse momentum $p_T$ of hadrons is defined
with respect to the weak current direction given by the vector
difference of the neutrino and muon momenta. \\ In Fig.~1a the
$p_T^2-$ distributions for $h^+$ and ${\pi}^-$ are plotted for the
whole sample of events. Both distributions are fitted to the form
$exp[-b{(p_T^2+m_{\pi}^2)}^{1/2}]$, with parameter $b(h^+) = 5.57
\pm 0.06$ and $b({\pi}^-) = 7.48 \pm$ 0.15 GeV$^{-1}$,
respectively. The latters are close to those extracted from the
neutrino interactions with a heavier composite target ($CF_3Br$)
at the same incident energies \cite{ref6}. The $p_T^2-$
distributions for the $B_S$ subsample are less steeper than for 
the $B_N$ subsample (both shown in Figs.~1b and 1c),
owing to an additional transverse momentum acquired by final 
hadrons in the intranuclear scattering processes. 
These distributions can be also
fitted to the same form resulting in $b_S(h^+) = 5.30\pm$ 0.08
and $b_S({\pi}^-) = 7.22\pm$ 0.21 GeV$^{-1}$ for the 'cascade'
subsample which are by about 10-15\% smaller than those in the
'quasinucleon' subsample, $b_N(h^+) = 6.06\pm$ 0.10 and
$b_N({\pi^-}) = 7.99\pm$ 0.25 GeV$^{-1}$. Fig~2. shows the
$p_T^2-$ distributions for charged hadrons $(h^{\pm})$ produced in
the current quark and the target fragmentation regions (with
Feynman variable $x_F > 0$ and  $x_F < 0$, respectively). While at
$x_F > 0$ the parameter $b$ is almost the same for subsamples
$B_N$ and $B_S$, $b_N(x_F > 0) = 6.57 \pm 0.12$ (GeV$/c)^{-1}$ and
$b_S(x_F > 0) = 6.43 \pm 0.13$ (GeV$/c)^{-1}$, at $x_F < 0$
$b_S(x_F < 0) = 5.29 \pm 0.10$ (GeV$/c)^{-1}$ is by 16\% smaller
than $b_N(x_F < 0) = 6.13 \pm 0.15$ (GeV$/c)^{-1}$. The 
values of $b_N(x_F > 0)$ and $b_N(x_F < 0)$ for the 'quasinucleon'
subsample $B_N$ are consistent with those measured in ${\nu}p$
\cite{ref7} and ${\nu}D$ \cite{ref8} interactions at somewhat
higher $<W^2>$. \\ More informative (with respect to the nuclear
effects) are the values of $<p_T^2>$ collected in Table 1 for
several subsamples and various regions of $x_F$ and $W^2$. As
it is seen, the nuclear effects influence faintly the $<p_t^2>$
for ${\pi}^-$ mesons, as well as for $h^+$ with $x_F > 0$ both in
low ($4 < W^2 < 9$ GeV$^2$) and high ($9 < W^2 < 25$
GeV$^2$) regions of $W^2$. On the contrary, 
the $<p_T^2>$ for $h^+$ with $x_F < 0$
increases by ${\Delta}(p_T^2) = <p_T^2>_S - <p_T^2>_N = 0.061 \pm
0.013$ (GeV$/c)^2$ due to the secondary intranuclear interactions.
One should emphasize, that this rise is
more prominent at $W^2 <$ 9 GeV$^2$, where ${\Delta}(p_T^2) =
0.077 \pm$ 0.016 (GeV$/c)^2$, than at $W^2 >$ 9 GeV$^2$, where
${\Delta}(p_T^2) = 0.043 \pm 0.021$ (GeV$/c)^2$. This is in accordance
with the recent observations \cite{ref4,ref5,ref9} that the
effects of the secondary intranuclear interactions are weakening with
increasing of $W^2$.

\begin{table}[ht]
\begin{center}
\begin
{tabular}{|l c c c c|}
  \hline


&$<p_T^2>_N$& $<p_T^2>_S$&$<p_T^2>_D$&$<p_T^2>_A$\\ \hline 
\multicolumn{5}{|c|}{} \\
 &\multicolumn{4}{c|}{4 $< W^2 <$ 25
GeV$^2$}\\  \multicolumn{5}{|c|}{} \\ $h^+(x_F > 0)$
&0.180$\pm$0.006&0.190$\pm$0.007&0.182$\pm$0.007&0.187$\pm$0.005
\\ ${\pi}^-(x_F > 0)$&0.129$\pm$0.009&0.128$\pm$0.008&0.131$\pm$0.009 &0.129$\pm$0.006
  \\
\multicolumn{5}{|c|}{} \\
 $h^+(x_F < 0)$& 0.207$\pm$0.009&0.268$\pm$0.009&0.205$\pm$0.009
&0.252$\pm$0.007\\ ${\pi}^-(x_F < 0)$& 0.126$\pm$0.009&
0.141$\pm$0.008 &0.127$\pm$0.009&0.137$\pm$0.006
\\ \hline \multicolumn{5}{|c|}{} \\
 &\multicolumn{4}{c|}{4 $<
W^2 <$ 9 GeV$^2$}\\ \multicolumn{5}{|c|}{} \\ $h^+(x_F >
0)$&0.163$\pm$0.007&0.170$\pm$0.007&0.161$\pm$0.007&0.166$\pm$0.005\\
${\pi}^-(x_F >
0)$&0.101$\pm$0.009&0.119$\pm$0.010&0.104$\pm$0.009&0.113$\pm$0.007
\\
\multicolumn{5}{|c|}{} \\ $h^+(x_F <
0)$&0.189$\pm$0.011&0.266$\pm$0.011&0.186$\pm$0.011&0.245$\pm$0.008
\\ ${\pi}^-(x_F < 0)$&0.106$\pm$0.009&0.127$\pm$0.009&0.108$\pm$0.009&0.122$\pm$0.007
 \\ \hline
\multicolumn{5}{|c|}{} \\ &\multicolumn{4}{c|}{9 $< W^2 <$ 25
GeV$^2$}\\ \multicolumn{5}{|c|}{} \\
 $h^+(x_F > 0)$&0.205$\pm$0.012&0.217$\pm$0.013&0.214$\pm$0.013&0.216$\pm$0.009 \\
${\pi}^-(x_F >
0)$&0.154$\pm$0.015&0.134$\pm$0.013&0.156$\pm$0.016&0.144$\pm$0.010\\
 \multicolumn{5}{|c|}{}  \\
$h^+(x_F <
0)$&0.228$\pm$0.015&0.271$\pm$0.014&0.229$\pm$0.016&0.259$\pm$0.011\\
${\pi}^-(x_F <
0)$&0.149$\pm$0.018&0.157$\pm$0.019&0.151$\pm$0.018&0.155$\pm$0.011\\
\hline
\end{tabular}

\end{center}
\caption{The mean values of $<p_T^2>$ in (GeV$/c)^2$ of $h^+$ and
${\pi}^-$ for several subsamples and various regions of $x_F$
and $W^2$.}
\end{table}

\noindent Another consequence of the intranuclear interactions is
that the region of $x_F > 0$ turns out to be somewhat depleted,
while the region of $x_F < 0$ enriched for the subsample $B_S$
relative to the subsample $B_N$. This can be clearly seen from the data
on the mean multiplicities presented in Table 2. The 
depletion and enrichment effects can be characterized by the
ratios ${\rho}(x_F> 0) =~  <n(x_F > 0)>_S/<n(x_F > 0)>_N$ 
and ${\rho}(x_F < 0)  = 
<n(x_F < 0)>_S/<n(x_F < 0)>_N$, respectively. For ${\pi}^-$
mesons, the depletion effect is rather weak, ${\rho}^{-}(x_F > 0)
= 0.92 \pm 0.05$, while the multiplicity gain at $x_F < 0$ reaches
${\rho}^{-}(x_F < 0) = 1.66 \pm$ 0.11. Slightly larger effects are
observed for positively charged hadrons: ${\rho}^{+}(x_F > 0) =
0.83\pm$ 0.02 and ${\rho}^{+}(x_F < 0) = 1.80 \pm$ 0.08. Note,
that latter value can be somewhat influenced by the contamination
from the non-identified recoil protons emitted during the
secondary interaction processes. The lower limit of the mean
multiplicity of these protons, evaluated from the identification efficiency
for protons with $0.6 < p_p <$ 0.85 GeV$/c$ (almost all having
$x_F <$ 0), turns out to be about 5\% of $<{n_{h^+}(x_F < 0)>}_S$.
The data of Table 2 also indicate that the depletion and
enrichment effects depend on $W$ only slightly.

\begin{table}[h]
\begin{center}
\begin
{tabular}{|l c c c c|}
  \hline


&$<n>_N$& $<n>_S$&$<n>_D$&$<n>_A$
\\ \hline
\multicolumn{5}{|c|}{} \\ &\multicolumn{4}{c|}
 {4 $< W^2 <$ 25 GeV$^2$}\\
 \multicolumn{5}{|c|}{} \\ $h^+(x_F > 0)$&1.441$\pm$0.023
  &1.198$\pm$0.023&1.402$\pm$0.023&1.279$\pm$0.017 \\
${\pi}^-(x_F > 0)
$&0.449$\pm$0.019&0.412$\pm$0.017&0.474$\pm$0.019&0.437$\pm$0.013
\\
 \multicolumn{5}{|c|}{} \\
$h^+(x_F <
0)$&0.725$\pm$0.023&1.303$\pm$0.036&0.677$\pm$0.023&1.052$\pm$0.024
\\ ${\pi}^-(x_F < 0)$&0.325$\pm$0.017&0.541$\pm$0.022&
0.339$\pm$0.018 &0.460$\pm$0.015 \\ \hline
 \multicolumn{5}{|c|}{} \\
 &\multicolumn{4}{c|}{4 $< W^2 <$ 9 GeV$^2$}\\
\multicolumn{5}{|c|}{} \\ $h^+(x_F >
0)$&1.349$\pm$0.026&1.106$\pm$0.027&1.319$\pm$0.026&1.193$\pm$0.019\\
${\pi}^-(x_F >
0)$&0.325$\pm$0.019&0.309$\pm$0.018&0.360$\pm$0.021&0.329$\pm$0.014
\\  \multicolumn{5}{|c|}{}  \\
$h^+(x_F <
0)$&0.632$\pm$0.027&1.172$\pm$0.040&0.589$\pm$0.027&0.935$\pm$0.027
\\
${\pi}^-(x_F <
0)$&0.277$\pm$0.019&0.480$\pm$0.024&0.295$\pm$0.020&0.405$\pm$0.016
 \\ \hline
\multicolumn{5}{|c|}{} \\ &\multicolumn{4}{c|}{9 $< W^2 <$ 25
GeV$^2$}\\ \multicolumn{5}{|c|}{} \\
 $h^+(x_F > 0)$&1.596$\pm$0.042&1.349$\pm$0.042&1.545$\pm$0.042&1.425$\pm$0.030 \\
${\pi}^-(x_F >
0)$&0.660$\pm$0.036&0.579$\pm$0.032&0.671$\pm$0.036&0.615$\pm$0.024\\
\multicolumn{5}{|c|}{} \\ $h^+(x_F <
0)$&0.883$\pm$0.043&1.516$\pm$0.066&0.829$\pm$0.043&1.248$\pm$0.044\\
${\pi}^-(x_F <
0)$&0.408$\pm$0.032&0.640$\pm$0.041&0.417$\pm$0.033&0.553$\pm$0.028
\\ \hline
\end{tabular}
\end{center}

\caption{The mean multiplicities of $h^+$ and ${\pi}^-$ for
several subsamples and various regions of $x_F$ and $ W^2$.}
\end{table}

\noindent To compare our data with the results of
other experiments, as well as to extract quantitative 
characteristics of the secondary intranuclear interactions, 
the data on $<p_T^2>$, corresponding to
${\nu}D$ and ${\nu}A$ interactions, $<p_T^2>_D$ and $<p_T^2>_A$,
respectively, are presented in Table 1. The values of
 $<p_T^2>_D$ and $<p_T^2>_A$ are defined as
\begin{equation}
<p_T^2>_D = \frac{0.6 N_p}{N_D}\, <p_T^2>_p + \frac{N_n}{N_D}\,
<p_T^2>_n \, \, ,
\end{equation}
\begin{equation}
<p_T^2>_A = \frac{N_D}{N_A}\, <p_T^2>_D + \frac{N_S}{N_A}\,
<p_T^2>_S \, \, .
\end{equation}

\noindent Here $N_D = N_n + 0.6 N_p$, $N_A = N_S + N_D$ and $
<p_T^2>_p$  and $<p_T^2>_n$ are the values of $<p_T^2>$ for
subsamples $B_p$ and $B_n$, respectively. \\ The average
multiplicities $<n>_A$ and $<n>_D$ are defined similarly (see Table 2). 
The measured values of ${<p_T^2>}_A - {<p_T^2>}_D$ and $<n>_A - <n>_D$
will be compared with theoretical predictions in Section 4.

\noindent The dependence of the mean value of $<p_T^2>$ of charged
hadrons (combined $h^+$ and ${\pi}^-$) on the DIS kinematical
variables $W^2$ and ${\nu}$ is presented in Figs.~3 and 4. \\ It is
seen from Fig.~3a, that $<p_T^2>$ for particles with $x_F
>0$ increases with $W^2$, as it was observed in earlier
investigations with neutrino and muon beams
\cite{ref8,ref10,ref11,ref12,ref13,ref14}. At the considered range
of $W^2$, this rise is essentially caused by increase in the available 
phase space. With increasing of $W^2$, the QCD effects are predicted
\cite{ref15} to play more and more significant role in the rise of
$<p_T^2>$. The data show (Figs.~3a and 3c), that the nuclear
effects practically do not influence $<p_T^2>$ at $x_F
>0$. On the contrary, they cause a significant increase of $<p_T^2>$
at $x_F <0$ in the region of $W^2 <$ 15 GeV$^2$, where the
difference ${<p_T^2>}_A - {<p_T^2>}_D$ is about 0.04 (GeV$/c)^2$ and
practically independent of $W^2$.
\\ The dependence  of $<p_T^2>$ on ${\nu}$ (see Fig.~4) 
reveals analogous significant nuclear effects for particles
with $x_F <0$ in the region of ${\nu}<$ 9 GeV, where the
difference ${<p_T^2>}_A - {<p_T^2>}_D$ is around (0.03-0.05)
(GeV$/c)^2$. At ${\nu}$ higher than 9 GeV, the nuclear effects
on $<p_T^2>$ are negligible. As for particles with $x_F >$0, the
effects of secondary intranuclear interactions practically do not
influence $<p_T^2>$.
\\ The dependence of $<p_T^2>$ for charged hadrons 
on their kinematical variables is presented in Figs.~5 and 6.\\ 
The dependence of $<p_T^2>$ on $x_F$ (Fig.~5a) has a typical 'seagull'
form (cf., for example, \cite{ref6,ref11}), both for $B_N$ and
$B_S$ subsamples. For the latter, a small part (about 3\%) of
positively charged hadrons occupies the region of $x_F <$ -1,
kinematically forbidden for reactions on a free nucleon (the so
called 'cumulative' region), where $<p_T^2>_S$ reaches a rather
high value of $<p_T^2>_S \sim$ 0.6 (GeV$c)^2$. About
third of these cumulative hadrons are estimated to be
non-identified protons with $0.6 < p_p <$ 0.85 GeV$/c$ and
$<p_T^2>$ = 0.51$\pm$0.06 (GeV$/c)^2$. It is seen from Fig.~5b,
that the influence of nuclear effects is significant only 
at $x_F <$ -0.6, where the difference
${<p_T^2>}_A - {<p_T^2>}_D$ is about (0.07$\pm$0.03) (GeV$/c)^2$.
\\ Fig.~6 shows the dependences of $<p_T^2>_N$, $<p_T^2>_S$ and
$<p_T^2>_A - <p_T^2>_D$ on the variable $z = E_h/{\nu}$, the
fraction of the current quark  energy carried by the hadron. The
rise of $<p_T^2>_N$ with $z$ for forward hadrons (Fig.~6c),
observed earlier in \cite{ref10,ref11,ref13}, is mainly caused by
the intrinsic transverse momentum of the current quark inside the
nucleon \cite{ref16} (see Section 4 for details). It is seen from
Fig.~6b, that the secondary intranuclear interactions induce a
significant difference of $<p_T^2>_A - <p_T^2>_D$ almost in the
whole range of $z$. The rize of $<p_T^2>_A - <p_T^2>_D$ at $0 < z
< 0.3$ is contributed by particles with $x_F < 0$ (cf. Fig.~6f),
while its fall at $z > 0.3$ is caused by the increasing
contribution of forward particles for which the difference
$<p_T^2>_A - <p_T^2>_D$ is rather small (Fig.~6d).

\section{Discussion}

The data on ${<p_T^2>}_N$ of hadrons with $x_F >$0 in the 
subsample $B_N$ were checked to be consistent with
the conventional picture of the quark string fragmentation (see e.g.
\cite{ref16} for review). According to the latter, the $z$-
dependence of $<p_T^2>_N$ for leading hadrons (containing the
current quark) can be parameterized as \cite{ref17}:
\begin{equation}
<p_T^2>_N = <p_T^2>_{Frag} + z^2<k_T^2> + <p_T^2>_{QCD} \, \, ,
\end{equation}
\noindent where  $<p_T^2>_{Frag}$ is the contribution from the
fragmentation process, $k_T$ is the primordial transverse momentum
of the current quark inside the nucleon, while $<p_T^2>_{QCD}$ is
the contribution of QCD effects (involving hard gluon emission and
$q \bar{q}$ production). At  $W^2$ available in this experiment
($W^2 <$ 25 GeV$^2$), the latter term can be approximately
parameterized as \cite{ref18}
\begin{equation}
<p_T^2>_{QCD} = a (W^2 - W_0^2) \, \, ,
\end{equation}

\noindent with $a = 3.5 \cdot 10^{-3}$ (GeV$/c)^2$ and $W_0^2$ = 2
GeV$^2$. \\ Fig.~7 shows the $z^2$- dependence for $<p_T^2>$ of
positively charged hadrons in the current fragmentation region for
two intervals of $W^2$, $W^2 <$ 9 GeV$^2$ and $W^2 >$ 9 GeV$^2$
(containing approximately equal statistics). The data for
fast hadrons (with $z^2 >$ 0.16), containing with a
high probability the current quark, are fitted to dependence (3).
The QCD term in (3) was fixed according to (4).
At $z^2 >$ 0.16 the mean value of $<W^2>$ in
present experiment is practically independent of $z^2$: $<W^2>
\approx$ 6 GeV$^2$ leading to $<p_T^2>_{QCD}$ = 0.014 (GeV$/c)^2$
for the region $W^2 <$ 9 GeV$^2$, and $<W^2> \approx$ 14 GeV$^2$
leading to $<p_T^2>_{QCD}$ = 0.042 (GeV$/c)^2$ for the region $W^2
>$ 9 GeV$^2$. The fit results are plotted in Figs.~7a and 7b.
The fitted values of $<p_T^2>_{Frag}$ and $<k_T^2>$ turn out to be
independent of $W^2$ within statistical uncertainties: 
$<p_T^2>_{Frag} = 0.17 \pm 0.03$ (GeV$/c)^2$, $<k_T^2> = 
0.23 \pm 0.10$ (GeV$/c)^2$ at
$W^2 <$ 9 GeV$^2$, and $<p_T^2>_{Frag} = 0.22 \pm 0.07$
(GeV$/c)^2$, $<k_T^2> = 0.30 \pm 0.19$ (GeV$/c)^2$ at $W^2 >$ 9
GeV$^2$. \\ The quoted values of $<k_T^2>$ are consistent with
those extracted from the data on $\nu$p \cite{ref19} and $\mu$p
\cite{ref20,ref21} deep inelastic scattering at higher $W^2$ (16
$< W^2 <$ 400 GeV$^2$). The values of  $<p_T^2>_{Frag}$ are also
consistent with that estimated in \cite{ref19}, however somewhat
underestimate the value of  $<p_T^2>_{Frag} = 0.274 \pm 0.059$
extracted from $e^+e^-$ annihilation at the LEP energies
\cite{ref22}. \\ In Fig.~7c the difference ${<p_T^2>}_A -
{<p_T^2>}_D$ versus $z^2$ is plotted. The data at low $W^2$, $W^2 <$ 9
GeV$^2$, indicate, that the additional transverse momentum,
acquired by forward hadrons (with $x_F >$ 0) due to the
intranuclear interactions, slightly increases with $z$, while at
larger $W^2 >$ 9 GeV$^2$ no significant nuclear effects are
observed.

\noindent Below an ~attempt is ~undertaken to describe the
obtained experimental ~data on ~differences $<n>_A -<n>_D$ and
$<p_T^2>_A - <p_T^2>_D$, which characterize the strength of
nuclear effects, with the help of a simple model incorporating
the secondary intranuclear interactions of produced pions. We
assume that the formation lenght $l_{\pi}$ of pions is determined
\cite{ref23} in the framework of the Lund fragmentation model
\cite{ref24}:
\begin{equation}
l_{\pi} = {\nu}\, z [\frac{ln(1/z^2) -1 + z^2}{1 - z^2}]\, /k \, ,
\end{equation}
\noindent where $k \approx$ 1 GeV/fm is the quark string tension.
The expression (5) has a maximum at $z \approx$ 0.3 and behaves as
$l_{\pi} \approx 2{\nu}\,z \ln(0.61/z) /k$ at $z <$ 0.2 and
approximately (with an accuracy better than 20\%) as $l_{\pi}
\approx {\nu}\, (1-z)\, /k$ at $z >$ 0.5. The latter behaviour
(predicted also in \cite{ref25}) was found to be consistent with
recent experimental data \cite{ref4,ref5,ref26}. We assume, that a
pion can interact in the nucleus, starting from the distance
$l_{\pi}$ from the ${\nu} N$ scattering point.\\ The hadrons
(predominantly pions) produced in ${\nu}\, N$ 
interactions at $x_F>$ 0 carry relatively large momenta, the
mean values  $\bar {p}_{\pi}(x_F >$ 0) of which are presented in 
the Appendix (Table A1) for two regions of $W^2$, 
$W^2 <$ 9 GeV$^2$  and  $W^2 >$ 9 GeV$^2$.
We assume, that these pions undergo intranuclear inelastic 
interactions with pion-nucleon cross section 
${\sigma}^{in}_{{\pi}N}$ (averaged over
protons and neutrons of the target nuclei). The estimated cross
sections are around ${\sigma}^{in}_{{\pi}N} \approx$ 21-24 mb
\cite{ref27,ref28}, depending on $\bar {p}_{\pi}(x_F >$ 0). 
With these cross sections, 
the probabilities  $w_{in}$ of secondary interactions
averaged over $l_{\pi}$ and nuclei of the propane-freon mixture
are calculated (see \cite{ref4,ref5} for details). The values of
$w_{in}$ are given in Table A1.
The average multiplicities $\bar{n} ({\pi}\rightarrow {\pi}^{'})$
of secondary ${\pi}^+$ and ${\pi}^-$ mesons in reactions ${\pi}\,
N \rightarrow {\pi}\, X$ at given average momenta $\bar {p}_{\pi}(x_F >$
0) are estimated using the partial ${\pi}\, N$ cross
sections \cite{ref27} (see Table A1).
Further one may assume, that a fraction $\beta$ of secondary
pions occupies the region of $x_F <$ 0, while the remaining
fraction $(1-\beta)$ stays in the region of $x_F >$ 0. The
parameter $\beta$ (a free parameter of the model) is expected to differ
for like-sign pions, i.e. for ${\pi}^+({\pi}^-)$ mesons in
${\pi}^+({\pi}^-)$- induced reactions and for unlike-sign pions,
i.e. for ${\pi}^-({\pi}^+)$ mesons in ${\pi}^+({\pi}^-)$- induced
reactions. For unlike-sign pions ${\beta}_u \sim$ 1 is expected, 
because they acquire a small fraction of the parent pion
energy and hence occupy predominantly the region of $x_F <$ 0. A
value of ${\beta}_u$ = 0.9$\pm$0.1 is chosen as a result. The parameter
${\beta}_l$ for like-sign pions has to be smaller than for unlike-sign
pions, ${\beta}_l < {\beta}_u$,
due to the leading hadron effect. A value of ${\beta}_l$ =
0.67$\pm$0.10 is chosen to fit the data on measured quantity
$<n_{{\pi}^{\pm}}(x_F> 0)>_A - <n_{{\pi}^{\pm}}(x_F > 0)>_D$ 
characterizing the attenuation
of the ${\pi}^{\pm}$ yield in the forward hemisphere of ${\nu}A$
interactions as compared to that in ${\nu}D$ interactions.\\
The intranuclear interactions of produced ${\pi}^0$ mesons 
should be taken into account also. 
It is expected that all relevant quantities
concerning ${\pi}^0$ are in between those for ${\pi}^+$ and
${\pi}^-$ (see \cite{ref29} and references therein). For simplcity
they are taken as averages of those for ${\pi}^+$ and
${\pi}^-$ mesons, with a maximal uncertainty. For example, the
parameter ${\beta}_0$ for ${\pi}^0$- induced reactions, is taken
to be ${\beta}_0 = 0.5({\beta}_l + {\beta}_u)$ with uncertainty
${\Delta \beta}_0 = \pm0.5 (\beta_l -\beta_u)$.
\\With the above quantities, characterizing the secondary
inelastic interactions, one can write down expressions (given by
eqs. A1 and A2 in the Appendix) for the predicted multiplicity
differences ${<n_{\pi}>}_A - <n_{\pi}>_D$ of pions produced in
${\nu}A$ and ${\nu}D$ interactions at $x_F >$ 0 and $x_F <$ 0. A
comparison with the experimental data is given in Table 3.

\begin{table}[h]
\begin{center}
\begin
{tabular}{|c|c|c|c|}
  \hline



 range&particle&\multicolumn{2}{c|}{$<n>_A-<n>_D$}
  \\ \cline{3-4}
of $W^2$(GeV$^2)$&type&measured&calculated \\ \hline
  &$h^+(x_F > 0)$
&-0.123$\pm$0.019&-0.136$\pm$0.045\\ &${\pi}^-(x_F >
0)$&-0.037$\pm$0.015&-0.008$\pm$0.032 \\ 4$< W^2 <$9&
 $h^+(x_F < 0)$& 0.375$\pm$0.025&0.407$\pm$0.084 \\
& ${\pi}^+(x_F < 0)$&$-$& 0.318$\pm$0.082 \\ &${\pi}^-(x_F <
0)$&0.121$\pm$0.017&0.241$\pm$0.069
\\ \hline 
 &$h^+(x_F >
0)$&-0.119$\pm$0.036&-0.059$\pm$0.041 \\ &${\pi}^-(x_F >
0)$&-0.056$\pm$0.029&-0.006$\pm$0.039 \\ 
9$< W^2 <$25&$h^+(x_F < 0)$&0.418$\pm$0.048&0.284$\pm$0.051
\\ &
${\pi}^+(x_F < 0)$&$-$&0.234$\pm$0.050 \\ & ${\pi}^-(x_F
<0)$&0.136$\pm$0.032&0.186$\pm$0.044 \\ \hline
\end{tabular}

\end{center}
\caption{The measured and predicted  differences {$<n>_A-<n>_D$}
at $x_F >$ 0 and $x_F <$ 0.}
\end{table}

\noindent A reasonable consistency with the data at
$x_F >$ 0 is observed. 
Particularly, the model predicts
a stronger depletion for the yield of $h^+$ than for ${\pi}^-$ in
the forward hemisphere in agreement with the data. 
The data description at $x_F <$ 0 is
worse. The model overestimates the enhancement of the
${\pi}^-$ yield at $W^2 <$ 9 GeV$^2$ by factor two,
but is in agreement with the data at $W^2 >$ 9 GeV$^2$ within
experimental uncertainties. 
On the other hand, the
predicted values~ of $<n_{{\pi}^+}(x_F < 0)>_A~
-~<n_{{\pi}^+}(x_F < 0)>_D$ are smaller than the measured values
for $h^+$, especially at $W^2 >$ 9 GeV$^2$.

\noindent As it was already mentioned, 
the positively charged hadrons with $x_F <$ 0 
in the subsample $B_S$ (and $B_A$) include 
non-identified protons with $p_p >$ 0.6 GeV$/c$. The latters
originate from secondary interactions of produced pions, both from
inelastic interactions (of pions with $x_F >$ 0) and elastic
${\pi}p$ scattering of pions with $x_F >$ 0 and $x_F <$ 0 
(average momenta are given in Table A1).
The probability of the intranuclear scattering
${\pi}p \rightarrow {\pi}p$ convoluted with the probability of
non-identification of the recoil proton is estimated using the
differential cross section of ${\pi}^{\pm}p \rightarrow
{\pi}^{\pm}p$ \cite{ref28}. The corresponding expressions for the
mean multiplicity $<n_{p}^{nid}>_{el}$ of non-identified recoil
protons are given in the Appendix (eqs. A3 and A4). The calculated
value of $<n_{p}^{nid}>_{el}$, taking into account the
contribution from ${\pi}^0p \rightarrow {\pi}^0p$\ also, turns out to
be 0.042$\pm$0.011 at $W^2 <$ 9 GeV$^2$ and 0.023$\pm$0.006 at
$W^2
>$ 9 GeV$^2$. \\ The probability of inelastic reactions ${\pi}N
\rightarrow p\,X$ convoluted with the probability of
non-identification of the recoil proton can be estimated using the
differential cross section $d{\sigma}_{in}/dt$, where $t$ is the
squared four-momentum transferred to proton. As there are no data 
available for cross sections of inclusive reactions 
${\pi}N \rightarrow pX$ 
at the relevant momentum range of the projectile pion, the
data \cite{ref28} on exclusive channels were used. The data can be 
parametrized as $d{\sigma}_{in}^p/dt = a_1 {\exp(b_p^{'}t)} + a_2
{\exp(b_p^{''}t)}$ with $b_p^{'}$ = 5$\pm$2 (GeV$/c)^{-2}$ and
$b_p^{''}$ = 1.5$\pm$0.5 (GeV$/c)^{-2}$. The coefficients $a_1$
and $a_2$ are extracted from the compilations
\cite{ref27,ref28}. The corresponding expression for
$<n_p^{nid}>_{inel}$ is given in the Appendix (eq. A5). The
calculated value of $<n_p^{nid}>_{inel}$, taking into account
the contribution from ${\pi}^0\, N \rightarrow pX$ also, turns out to
be 0.047$\pm$0.020 at $W^2 <$ 9 GeV$^2$ and 0.027$\pm$0.010 at
$W^2 >$ 9 GeV$^2$. Using the summary multiplicity $<n_p^{nid}> =
<n_p^{nid}>_{el} + <n_p^{nid}>_{inel}$, the overwhelming fraction
of which refers to the region of $x_F <$ 0, one obtains
predictions for
\begin{eqnarray}
<n_{h^+}(x_F <0)>_A - <n_{h^+}(x_F <0)>_D = \nonumber  \\
<n_{{\pi}^+}(x_F < 0)>_A + <n_p^{nid}> - <n_{{\pi}^+}(x_F <0)>_D
\, ,
\end{eqnarray}
assuming $<n_{h^+}(x_F <0)>_D \approx <n_{{\pi}^+}(x_F <0)>_D$ for
${\nu}D$ interactions. The latter correction, as one may observe,
slightly improves the agreement with the data on
$<n_{h^+}(x_F <0)>_A - <n_{h^+}(x_F <0)>_D$. A significant
discrepancy, however, remains for the data at $W^2 >$ 9 GeV$^2$.
Nevertheless, the model reproduces the nuclear depletion 
and enhancement effects for the yield of $h^+$ and ${\pi}^-$ 
qualitatively and, in particular, predicts, in accordance with the
experimental observation, these effects to be more significant for
$h^+$ than for ${\pi}^-$.

\noindent A further application of the model concerns the
description of the data on $<p_T^2>$ presented in Table 1, namely,
on the difference ${<p_T^2>}_A - {<p_T^2>}_D$, being caused by
secondary intranuclear interactions of produced hadrons, both
elastic and inelastic. \\ The elastic ${\pi}N$ cross sections
${\sigma}_{el}(|t| > t_{min})$ are extracted from the data
\cite{ref28} on differential cross sections integrated over the
region of $|t| > t_{min} = 0.05$ (GeV$/c)^2$. The latter restriction 
is introduced in order to take into account the Pauli blocking
effect, i.e. the suppression of the intranuclear scattering with
small four-momenta transferred to the bound nucleon. The
probability $w_{el}$ of the elastic scattering (provided that the
pion did not undergo any inelastic interaction) is calculated at
average momenta $\bar {p_{\pi}}$ of ${\pi}^+$ and
${\pi}^-$ mesons with $x_F >$ 0 and $x_F <$ 0 (see Table A1). 
The average
${<p_T^2>}^{{\pi}N}_{el}$ acquired by the elastically scattered
pion (with respect to the 'projectile' one) varies from 0.09 to
0.26 (GeV$/c)^2$, depending on $\bar {p_{\pi}}$ and the pion sign.
The corresponding values with respect to the weak current direction,
${<p_T^2>}_{el}$, can be obtained as
\begin{equation}
{<p_T^2>}^{\pi}_{el} = {<p_T^2>}^{{\pi}N}_{el}(1- \frac{3}{2} \,
<{\sin}^2{\vartheta}_{\pi}>) +
p^2_{sc}{<{\sin}^2{\vartheta}_{\pi}>} \, \, ,
\end{equation}
\noindent where $p_{sc}$ is the momentum of the scattered pion,
while ${\vartheta}_{\pi}$ is the polar angle of the 'projectile'
pion with respect to the weak current axis. The estimated values of
${<p_T^2>}^{\pi}_{el}$ are given in Table A1. \\
As for inelastic interactions of pions with $x_F >$ 0, there are no 
data on inclusive characteristics of reactions ${\pi}^+N  \rightarrow
{\pi}^{\pm}X$ and ${\pi}^-N  \rightarrow {\pi}^{\mp}X$ at the
relevant incident momenta. 
Therefore, ${<p_T^2({\pi} \rightarrow {\pi^{'}})>}_{in}$ 
is considered as a free parameter of the model (for
the each incident momentum and the reaction type). The general
expectation is that it increases with the
incident momentum and is larger for like-sign pions than for
unlike-sign ones \cite{ref30}. In addition, the final pions occupying
the region of $x_F <$ 0 are expected to have on an average larger
${<p_T^2({\pi} \rightarrow {\pi^{'}})>}^{b}_{in}$ than those with 
$x_F >$ 0, ${<p_T^2({\pi} \rightarrow {\pi^{'}})>}^{f}_{in}$, due to the
Lorentz transformation from the $\pi\, N$ c.m.s. to the c.m.s. of
the hadronic system. The chosen values of 
${<p_T^2({\pi} \rightarrow {\pi^{'}})>}^{f}_{in}$ and
${<p_T^2({\pi} \rightarrow {\pi^{'}})>}^{b}_{in}$ with
respect to the weak current axis are given in the Appendix (Table A2). 
One should note, that the maximal value of 
${<p_T^2({\pi^+} \rightarrow {\pi^+})>}^{b}_{in}$ = 0.3 (GeV$/c)^2$ 
refers to the like-sign pion with $x_F <$ 0 in reaction ${\pi}^+N
\rightarrow {\pi}^{+}(x_F <0)X$, while the minimal value of
${<p_T^2({\pi^-} \rightarrow {\pi^+})>}^{f}_{in}$ = 0.1 (GeV$/c)^2$ 
refers to the unlike-sign pion
with $x_F >$ 0 in reaction ${\pi}^-N \rightarrow {\pi}^{+}(x_F
>0)X$. \\ Using all above defined quantities, and taking into account the
contribution from ${\pi}^0$- induced reactions ${\pi}^0N
\rightarrow {\pi}^{\pm}X$ also, predictions for $<p_T^2>_A$ can be
obtained (see eqs. A6 and A7 in the Appendix). The predicted and 
measured values for the difference ${<p_T^2>}_A - {<p_T^2>}_D$ 
are presented in Table 4. The model reproduces rather small
values of this difference at $x_F >$ 0, as well as the data for
${\pi}^-$ mesons with with $x_F <$ 0. A reasonable agreement with
the data on ${<p_T^2(x_F <0)>}^{+} _A- {<p_T^2(x_F< 0)>}^{+}_D$ is
obtained when taking into account the contribution from
non-identified protons (see eq. A7 of the Appendix). 
The quoted errors in Table 4 for predicted values include only the
uncertainty in parameters $\beta_u$, $\beta_l$ and $\beta_0$ and
in the contribution from $\pi^0$- induced reactions, but do not
that in the chosen values of $<p_T^2>_{in}$ (quoted in Table A2 of
the Appendix).

\begin{table}[h]
\begin{center}
\begin{tabular}{|c|c|c|c|}
  \hline



 range&particle&\multicolumn{2}{c|}{{$<p_T^2>_A-<p_T^2>_D$}(GeV$/c)^2$}
  \\ \cline{3-4}
of $W^2$(GeV$^2)$&type&measured&calculated \\ \hline
  &$h^+(x_F > 0)$
&0.005$\pm$0.006&0.003$\pm$0.008\\ &${\pi}^-(x_F >

0)$&0.009$\pm$0.008&0.001$\pm$0.011 \\ 4$< W^2 <$9&
 $h^+(x_F < 0)$& 0.059$\pm$0.009&0.032$\pm$0.019 \\
& ${\pi}^+(x_F < 0)$&$-$& 0.020$\pm$0.020 \\ &${\pi}^-(x_F <
0)$&0.014$\pm$0.008&0.036$\pm$0.024
\\ \hline 
 &$h^+(x_F >
0)$&0.002$\pm$0.011&0.000$\pm$0.012 \\ &${\pi}^-(x_F >
0)$&-0.012$\pm$0.013&-0.002$\pm$0.015 \\ 
9$< W^2 <$25&$h^+(x_F < 0)$&0.030$\pm$0.013&0.010$\pm$0.016
\\ &
${\pi}^+(x_F < 0)$&$-$&0.006$\pm$0.016 \\ & ${\pi}^-(x_F
<0)$&0.004$\pm$0.008&0.016$\pm$0.019 \\ \hline
\end{tabular}

\end{center}
\caption{The measured and predicted differences
$<p_T^2>_A-<p_T^2>_D$ at $x_F >$ 0 and $x_F <$ 0.}
\end{table}

More definite predictions for $<p_T^2>_A-<p_T^2>_D$ can be
obtained for leading hadrons (the experimental data for which are
plotted in Fig.~7c). For the subsample $B_A$, these hadrons can be
subdivided into two fractions. The first one consists of hadrons
which escape the nucleus without any interaction (with a
probability $w_e$) and for which the value of $<p_T^2>_D$ is preserved.
The second fraction consists of hadrons which have undergone only
an elastic scattering (with a probability $w_{el}^f$) with a
moderate momentum transfer $|t| < t_{max}$ (in order to be
survived as a large -$z$ particle) and which acquire on an average
$<p_T^2>_{el}$ with respect to the weak current axis. The values of
$|t|$ in the range of $|t| < t_{max}$ correspond to the scattering
angle $\vartheta_{\pi}^* < 90^{\circ}$ in the ${\pi}N$ c.m.s. at
which the pion keeps on an average more than 90\% of its energy.
The value of $<p_T^2>_{el}$ is extracted using the data
\cite{ref28} on $d{\sigma}_{el}^{\pi N}/dt$ in the range of
$t_{min} < |t| < t_{max}$ and the relation (7) to transform it
with respect to the weak current axis. \\ The relative weights of these
two fractions of the leading hadrons are $w_e/(w_e + w_{el}^f)$ and
$w_{el}^f/(w_e + w_{el}^f)$, respectively. As a result,
\begin{equation}
<p_T^2>_A - <p_T^2>_D = \frac{w_{el}^f}{w_e +
w_{el}^f}(<p_T^2>_{el} - <p_T^2>_D) \, \, .
\end{equation}
The predicted values of (8) are compared with the data plotted in
Fig.~7a for positively charged leading hadrons with 0.16 $< z^2 <$
0.3 and $z^2 >$ 0.3 (at $W^2 <$ 9 GeV$^2$). In these intervals,
the particles are characterized by the following measured and
calculated quantities: $\bar {p}_{h^+}$ = 2.0 and 2.9 GeV$/c$,
$<p_T^2>_D$ = 0.227$\pm$0.015 and 0.297$\pm$0.024 (GeV$/c)^2$;
$\bar{l}_h$ = 1.9 and 1.1 fm, ${\sigma}_{el}^f(t_{min} < |t| <
t_{max})$ = 5.3 and 4.2 mb, ${\sigma}_{tot}^{'} =
{\sigma}_{in}^{{\pi}N} + {\sigma}_{el}^{'}(|t| > t_{min})$ = 28.5
and 27.7 mb, $w_{el}^f$ = 0.030 and 0.029, $w_e$ = 0.754 and
0.668, $<p_T^2>_{el}$ = 0.411 and 0.489 (GeV$/c)^2$, respectively.
With these values, one obtains for the difference $<p_T^2>_A -
<p_T^2>_D$ = 0.007$\pm$0.001 and 0.008$\pm$0.001 (GeV$/c)^2$ at
0.16 $< z^2 <$ 0.3 and $z^2 >$ 0.3, respectively. \\ The
corresponding quantities for $h^+$ with $z^2 >$ 0.2 (at $W^2 >$ 9
GeV$^2$) are: $\bar {p}_{h^+}$ = 6.1 GeV$/c$, $<p_T^2>_D$ =
0.401$\pm$0.042 (GeV$/c)^2$; $\bar{l}_h$ = 3.3 fm,
${\sigma}_{el}^f(t_{min} < |t| < t_{max})$ = 3.5 mb,
${\sigma}_{tot}^{'}$ = 20.8 mb, $w_{el}^f$ = 0.017, $w_e$ = 0.832
and  $<p_T^2>_{el}$ = 0.68 (GeV$/c)^2$, leading to $<p_T^2>_A -
<p_T^2>_D$ = 0.006$\pm$0.001 (GeV$/c)^2$. The predicted values of
$<p_T^2>_A - <p_T^2>_D$ for leading particles are rather small 
in agreement with the data and do not contradict the trend of the
latters with variation of $z^2$ and $W^2$, as it can be seen from
Fig.~7c.

\noindent Finally, one needs to note, that although the applied model 
reproduces
the majority of the experimental data (presented in Tables 3 and 4
and Fig.~7c), it is rather crude and uses several simplified
assumptions which should be summarized:\, i) the
calculations concerning the secondary intranuclear interactions
are performed for fixed average momenta of $\pi^+$ and $\pi^-$
mesons, $\bar{p}_{\pi^{\pm}}(x_F > 0)$ and $\bar{p}_{\pi^{\pm}}(x_F < 0)$,
instead of more extensive calculations averaged over the
momentum spectra; \, ii) the second-order effects of two or more
intranuclear collisions of a pion are neglected; \, iii) the model
does not incorporate the production of hadronic resonances, in
particular, $\rho$ mesons (composing about 10\% of charged pions
\cite{ref31,ref32}), with a proper space-time structure of their
formation, intranuclear interactions and decay.

\section{Summary}

New experimental data concerning the influence of the nuclear
medium on the transverse momentum of neutrinoproduced hadrons are
presented. \\ The  $p_T^2$- distribution of
hadrons (both positively and negatively charged) is less steeper
in 'cascading' (and 'nuclear') than in 'quasinucleon' (and
'quasideuteron') interactions. The influence of the nuclear medium
on the dependence of $<p_T^2>$ on kinematical variables of the DIS
and of final hadrons is studied. The nuclear
effects, leading to an enhancement of $<p_T^2>$, are more
prominent for the following ranges of variables: \\
 - for $x_F <$ 0 at $W^2 <$ 15 GeV$^2$ or $\nu <$ 9 GeV, while no
significant enhancement of $<p_T^2>$ is observed at higher $W^2$
or $\nu$;
\\ - for $x_F <$ -0.6, while at $x_F >$ -0.6 the manifestation of
nuclear effects is faint;
\\ - practically for the whole range of $z$.

\noindent The observed $z^2$- dependence of $<p_T^2>_N$ for fast
hadrons in the 'quasinucleon' subsample follows the conventional
picture of the quark string fragmentation. The extracted
parameters governing the transverse momentum of produced hadrons,
$<p_T^2>_{Frag}$ = 0.19$\pm$0.03 (GeV/$c)^2$ and $<k_T^2>$ =
0.24$\pm$0.09 (GeV/$c)^2$ (estimated for the whole range of $4 <
W^2 <$ 25 GeV$^2$), are compatible with values obtained at higher
energies.

\noindent The experimental data on nuclear effects are
compared with predictions of a simple model incorporating the
secondary intranuclear interactions of produced hadrons. 
The model predicts a depletion of the
particle yield at $x_F >$ 0 and an enhancement of that at $x_F <$
0 (more pronounced for positively charged hadrons for both regions
of $x_F >$ 0 and $x_F <$ 0) in agreement with the data. 
The quantitative description of the
most part of the data on the hadron yields is also acceptable. The
model describes also the data on the difference $<p_T^2>_A -
<p_T^2>_D$ both for $h^+$ and ${\pi}^-$ with $x_F >$ 0 and $x_F <$
0, as well as for the leading particles with $z >$ 0.4. It has
been shown, that the model predictions for differences
$<n_{h^+}(x_F < 0)>_A - <n_{h^+}(x_F < 0)>_D$ and $<p_T^2(x_F
< 0)>^{+}_A - <p_T^2(x_F < 0)>^{+}_D$ for positively charged hadrons
with $x_F <$ 0 turn out to be in better agreement with the data, 
when taking into account the contribution from non-identified protons.

\noindent {\bf{Acknowledgement:}} The activity of one of the
authors (Zh.K.) is supported by Cooperation Agreement between DESY
and YerPhI signed on December 6, 2002.

\newpage


\section*{Appendix}
\setcounter{equation}{0} \setcounter{table}{0}
\renewcommand{\theequation}{A\arabic{equation}}
\renewcommand{\thetable}{A\arabic{table}}

The mean multiplicity of $\pi^+$ mesons with $x_F
>$ 0 and $x_F <$ 0 in $\nu A $ interactions can be
approximately expressed, taking into account the intranuclear
inelastic interactions of produced pions, as:
\begin{eqnarray}
<n_{{\pi}^+}(x_F > 0)>_A  = (1 - w_{in}^+) <n_{{\pi}^+}(x_F >
0)>_D  \nonumber
\\  + \sum_{i=1}^{3}w_{in}^i \, <n_{{\pi}^i}(x_F > 0)>_D \,
(1-{\beta}_i)\, \bar{n} \, ({\pi}^{i} \rightarrow {\pi}^{+})
\end{eqnarray}

\begin{eqnarray}
 <n_{{\pi}^+}(x_F < 0)>_A\, =\, <n_{{\pi}^+}(x_F < 0)>_D
\nonumber \\ + \sum_{i=1}^{3} w_{in}^i \, <n_{{\pi}^i}(x_F > 0)>_D
\, {\beta}_i \, \bar{n} \, ({\pi}^{i} \rightarrow {\pi}^{+}) \, .
\end{eqnarray}

\noindent where the index $i$ refers to the pion species,
$\pi^+\,, \pi^-\,, \pi^0$. The first term in (A1) corresponds to
$\pi^+$ mesons produced in $\nu N$ (or $\nu D$) interactions with
$x_F
>$ 0 and not suffered inelastic interactions. The last three terms
correspond to $\pi^+$ mesons produced in secondary inelastic
interactions $\pi^+ N \rightarrow \pi^+ X$, $\pi^- N \rightarrow
\pi^+ X$ and $\pi^0 N \rightarrow \pi^+ X$, having mean
multiplicities $\bar{n} (\pi^+ \rightarrow \pi^+)$, $\bar{n}
(\pi^- \rightarrow \pi^+)$ and $\bar{n} (\pi^0 \rightarrow
\pi^+)$, and occupying the region of $x_F >$ 0 with probabilities
$(1 - \beta_l)$, $(1 - \beta_u)$ and $(1 - \beta_0)$,
respectively. The first term in (A2) corresponds to $\pi^+$ mesons
with $x_F < 0$ in $\nu D$ interactions. The secondary inelastic
interactions of these mesons play a negligible role, because their
average momentum is small (see Table A1). We neglect also
the charge-exchange reactions $\pi^+ n \rightarrow \pi^0 p$ and
$\pi^0 p \rightarrow \pi^+ n$, the summary contribution of which
to the mean multiplicity of final $\pi^+$ mesons is expected to be
largely cancelled. The last three terms in (A2) correspond to
$\pi^+$ mesons produced in secondary inelastic interactions
(mentioned above) and occupying the region of $x_F <$ 0 with
probabilities $\beta_l, \, \beta_u$ and $\beta_0$ for $\pi^+, \,
\pi^-$ and $\pi^0$- induced reactions, respectively. Similar
expressions~ can be~ written down for the mean multiplicity
of~ $\pi^-$ mesons, $<n_{{\pi}^-}(x_F > 0)>_A$ and
$<n_{{\pi}^-}(x_F < 0)>_A$. The values of $w_{in}^+, \, w_{in}^-,
\, w_{in}^0$, $\bar{n} (\pi^+\rightarrow \pi^{\pm}), \,
\bar{n} (\pi^- \rightarrow \pi^{\pm})$
and $\bar{n} (\pi^0 \rightarrow \pi^{\pm})$ are given in Table A1, 
while $\beta_l, \, \beta_u,$ \ and $\beta_0$ are defined in Section 4. \\
Contributions to the multiplicity of positively charged hadrons, 
${<n_{h^+}(x_F < 0)>}_A$, comes from $<n_{{\pi}^+}(x_F < 0)>_A$ 
and non-identified recoil protons with $p_p >$ 0.6
GeV$/c$, the overwhelming part of which occupies the region of
$x_F <$ 0. The mean multiplicity of these protons can be expressed
as follows: \\ i) The contribution from $\pi p$ elastic scattering
of $\pi^+$, $\pi^-$ and $\pi^0$ mesons with $x_F >$ 0, which
scatter with probabilities $w_{el}^+(f)$, $w_{el}^-(f)$ and
$w_{el}^0(f)$ (see Table A1), respectively:
\begin{equation}
{<n_p^{nid}>}^{\prime}_{el} = \sum_{i=1}^{3} \, w^i_{el}\, (f)
<n_{\pi^i}(x_F > 0)>_D \, {<n_p^{nid}>}^{el}_{{\pi_f^i} p} \, ,
\end{equation}
\noindent where ${<n_p^{nid}>}^{el}_{{\pi_f^+} p}$ =
0.4,\, ${<n_p^{nid}>}^{el}_{{\pi_f^-} p}$ = 0.27, \,
${<n_p^{nid}>}^{el}_{{\pi_f^0} p}$ = 0.34 at $W^2 <$ 9 GeV$^2$,
resulting in ${<n_p^{nid}>}^{\prime}_{el}$ = 0.038, and
${<n_p^{nid}>}^{el}_{{\pi_f^+} p}$ = 0.13,
${<n_p^{nid}>}^{el}_{{\pi_f^-} p}$ = 0.19,
${<n_p^{nid}>}^{el}_{{\pi_f^0} p}$ = 0.16, resulting in
${<n_p^{nid}>}^{\prime}_{el}$ = 0.004 at $W^2 >$ 9 GeV$^2$. \\ 

ii) The contribution from $\pi p$ elastic scattering of $\pi^+$,
$\pi^-$, and $\pi^0$ mesons with $x_F <$ 0, which scatter with
probabilities $w_{el}^+(b)$, $w_{el}^-(b)$ and $w_{el}^0(b)$,
respectively (see Table A1):
\begin{equation}
{<n_p^{nid}>}^{{\prime \prime}}_{el} = \sum_{i=1}^{3} \,
w^i_{el}\, (b) <n_{\pi^i}(x_F < 0)>_D \,
{<n_p^{nid}>}^{el}_{{\pi_b^i} p} \, \, ,
\end{equation}
\noindent where  ${<n_p^{nid}>}^{el}_{{\pi_b^+} p}$ =
0.05,\, ${<n_p^{nid}>}^{el}_{{\pi_b^-} p}$ = 0.09, \,
${<n_p^{nid}>}^{el}_{{\pi_b^0} p}$ = 0.07 at $W^2 <$ 9 GeV$^2$,
resulting in ${<n_p^{nid}>}^{{\prime \prime}}_{el}$ = 0.004, 
and  ${<n_p^{nid}>}^{el}_{{\pi_b^+} p}$ = 0.22,
${<n_p^{nid}>}^{el}_{{\pi_b^-} p}$ = 0.27,
${<n_p^{nid}>}^{el}_{{\pi_b^0} p}$ = 0.24, resulting in
${<n_p^{nid}>}^{{\prime \prime}}_{el}$ = 0.018 at $W^2 >$ 9 GeV$^2$. \\ 

iii) The contribution from $\pi N$ inelastic interactions of
$\pi^+$, $\pi^-$ and $\pi^0$ mesons with $x_F >$ 0, which interact
with probabilities $w_{in}^+$, $w_{in}^-$ and $w_{in}^0$,
respectively (see Table A1):
\begin{equation}
{<n_p^{nid}>}_{inel} = \sum_{i=1}^{3} \, w^i_{in}\, <n_{\pi^i}(x_F
< 0)>_D \, {<n_p^{nid}>}^{in}_{{\pi^i} p} \, \, ,
\end{equation}
\noindent where ${<n_p^{nid}>}^{in}_{{\pi^+} N} \approx
{<n_p^{nid}>}^{in}_{{\pi^-} N} \approx {<n_p^{nid}>}^{in}_{{\pi^0}
N} \approx$ 0.16 at $W^2 < $9 GeV$^2$, resulting in
${<n_p^{nid}>}_{inel}$ = 0.047, and
${<n_p^{nid}>}^{in}_{{\pi^+} N} \approx
{<n_p^{nid}>}^{in}_{{\pi^-} N} \approx {<n_p^{nid}>}^{in}_{{\pi^0}
N} \approx$ 0.17, resulting in ${<n_p^{nid}>}_{inel}$ =0.027
at $W^2 >$9 GeV$^2$.

\noindent The  summary multiplicity  of non-identified protons
\,is \, $<n_p^{nid}>  = {<n_p^{nid}>}^{\prime}_{el}  +
{<n_p^{nid}>}^{{\prime \prime}}_{el} \\ + <n_p^{nid}>_{inel}$ =
0.089 at $W^2 <$ 9 GeV$^2$ and 0.049 at $W^2 >$ 9 GeV$^2$. These
values have to be added to (A2) to obtain ${<n_{h^+}(x_F < 0)>}_A
= {<n_{\pi^+}(x_F < 0)>}_A + <n_p^{nid}>$.

\noindent Different fractions of $\pi^+$ mesons in (A1)-(A2) or of
positively charged hadrons in (A1)-(A5) carry different 
values of $<p_T^2>$.
Hence, the mean value of $<p_T^2>_A$ has to be expressed as a
superposition of different terms. For example, the expressions for
$<p_T^2(x_F > 0)>^{+}_A$ and $<p_T^2(x_F < 0)>^{+}_A$ have the
following form:
\begin{equation}
<p_T^2(x_F > 0)>^{+}_A \, = \, \sum_{i=1}^{5}\, {\alpha}_{i}^{f}\,
<p_T^2>_i^f \, \,
\end{equation}
\begin{equation}
<p_T^2(x_F < 0)>^{+}_A \, = \, \sum_{i=1}^{14}\,
{\alpha}_{i}^{b}\, <p_T^2>_i^b \, \,
\end{equation}
\noindent where ${\alpha}_{i}^{f}\, ({\alpha}_{i}^{b})$ is the
normalized part of the $i$- th fraction of $h^+$~ occupying~ the
region of ~$x_F >$ 0 \, $(x_F <$ 0). \\ The first term in (A6)
corresponds to 'primary' $\pi^+$ mesons with $x_F >$ 0 produced in
$\nu N$- interactions, which do not suffer any secondary
interaction within the nucleus and hence keep the value of
$<p_T^2(x_F > 0)>^{+}_D$. The second term corresponds to $\pi^+$
mesons with $x_F >$ 0, which have suffered an elastic scattering
(but not any inelastic interaction) and acquired $<p_T^2(x_F >
0)>^{\pi^+}_{el}$ (given in Table A1). The last three terms correspond
to $\pi^+$ mesons which are produced in secondary inelastic
interactions of 'primary' $\pi^+\, ,\pi^-$ and $\pi^0$ and have
occupied the region of $x_F >$ 0, acquiring $<p_T^2({\pi^+
\rightarrow \pi^+})>_{in}^f$, $<p_T^2({\pi^- \rightarrow
\pi^+})>_{in}^f$ and $<p_T^2({\pi^0 \rightarrow \pi^+})>_{in}^f$,
respectively. The latters are free parameters of the model and are
presented in Table A2. \\ The first term in (A7) corresponds to
'primary' $\pi^+$ mesons with $x_F <$ 0, which do not suffer any
interaction and hence keep the value of $<p_T^2(x_F < 0)>^{+}_D$. 
The second term corresponds to 'primary' $\pi^+$ mesons
with $x_F <$ 0, which have suffered an elastic scattering and
acquired $<p_T^2(x_F < 0)>^{\pi^+}_{el}$ (given in Table A1). The
next three terms correspond to $\pi^+$ mesons which are produced
in secondary inelastic interactions of 'primary' $\pi^+\, , \pi^-$
and $\pi^0$ and have occupied the region of $x_F <$ 0, acquiring
$<p_T^2({\pi^+ \rightarrow \pi^+})>_{in}^b$, $<p_T^2({\pi^-
\rightarrow \pi^+})>_{in}^b$ and $<p_T^2({\pi^0 \rightarrow
\pi^+})>_{in}^b$, respectively. The latters are free parameters.
Their values adopted for this study are presented in Table A2. 
The next six terms concerns
non-identified protons (with $p_p >$ 0.6 GeV$/c$) originating from
the elastic $\pi p$ scattering of pions with $x_F >$ 0 and $x_F <$
0, as a result of which the proton acquires $<p_T^2({\pi^+
p})>_{el}^f$, $<p_T^2({\pi^- p})>_{el}^f$, $<p_T^2({\pi^0
p})>_{el}^f$ and $<p_T^2({\pi^+ p})>_{el}^b$, $<p_T^2({\pi^-
p})>_{el}^b$, $<p_T^2({\pi^0 p})>_{el}^b$, respectively. The
latters (quoted in Table A2) are extracted from the data on
$d{\sigma}_{el}^{\pi N}/dt$ convoluted with the probability
(depending on $p_p$) of the proton non-identification. The last
three terms in (A7) corresponds to non-identified protons,
originating from inelastic reactions $\pi N \rightarrow p X$, as a
result of which the recoil proton acquires $<p_T^2({\pi^+
\rightarrow p})>_{in}$, $<p_T^2({\pi^- \rightarrow p})>_{in}$ and
$<p_T^2({\pi^0 \rightarrow p})>_{in}$ (quoted in Table A2). Due to
the lack of data, the latters are assumed to be a simple average:
$<p_T^2({\pi \rightarrow p})>_{in} = ({<p_T^2({\pi \rightarrow
\pi^+})>_{in}^b + <p_T^2({\pi \rightarrow \pi^-})>_{in}^b + <p_T^2
(\pi p)>_{el}})/3$. \\ Similar equations can be written down for
$<p_T^2 (x_F > 0)>^{-}_A$  (including all corresponding terms of
(A6)) and  $<p_T^2 (x_F < 0)>^{-}_A$ (including the first five
corresponding terms of (A7)). The chosen values of the
corresponding free parameters $<p_T^2({\pi \rightarrow
\pi^-})>_{in}^f$ and $<p_T^2({\pi \rightarrow \pi^-})>_{in}^b$ \,
($\pi \equiv \pi^+, \pi^-, \pi^0$) are given in Table A2.

\begin{table}[h]
\begin{center}
\begin{tabular}{|l c c c c c c|} \hline
    
\multicolumn{7}{|c|}{} \\

'Incident'& $\bar{p}_{\pi}$&$w_{in}$&$\bar{n}({\pi}\rightarrow{\pi}^+)$ 
&$\bar{n}({\pi}\rightarrow{\pi}^-)$&$w_{el}$&${<p^2_T>}^{\pi}_{el}$ \\
~~pion&(GeV$/c$)& &&&&(GeV$/c)^2$ \\  \hline 
\multicolumn{7}{|c|}{} \\
\multicolumn{7}{|c|}{4 $< W^2 <$ 9  GeV$^2$}  \\
\multicolumn{7}{|c|}{} \\ 
${\pi}^+(x_F>0)$& 1.7 &0.20& 1.14&
0.42&0.035& 0.31  \\  
${\pi}^-(x_F>0)$&1.3&0.19&0.35&1.03&0.041&0.25 \\
\multicolumn{7}{|c|}{} \\
${\pi}^+(x_F <0)$&0.71 &$-$&$-$&$-$&0.14&0.18 \\
${\pi}^-(x_F < 0)$&0.54&$-$&$-$&$-$&0.16&0.16\\ \hline
\multicolumn{7}{|c|}{} \\
 \multicolumn{7}{|c|}{9 $< W^2 <$ 25
GeV$^2$}\\ 
\multicolumn{7}{|c|}{} \\ 
${\pi}^+(x_F>0)$& 2.9 &0.10& 1.31&
0.48&0.016& 0.41  \\
${\pi}^-(x_F>0)$&2.4&0.09&0.52&1.24&0.02&0.37 \\
\multicolumn{7}{|c|}{}  \\  
${\pi}^+(x_F <0)$&0.85&$-$ &$-$&$-$&0.07&0.22 \\
${\pi}^-(x_F < 0)$&0.68&$-$ &$-$&$-$&0.11&0.16\\ \hline
\end{tabular}
\end{center}

\caption{The mean momenta $\bar{p}_{\pi}$ of pions produced in
$\nu N$ interactions, the probabilities of their inelastic ($w_{in}$)
and elastic ($w_{el}$) intranuclear interactions, the mean 
multiplicities $\bar{n}({\pi \rightarrow \pi^{'}})$ of secondary
pions, and the mean squared transverse momentum $<p^2_T>^{\pi}_{el}$
(with respect to the weak current axis) acquired by pions
undergone elastic $\pi N$ scattering,  
at $W^2 < 9$ GeV$^2$ and $W^2 > 9$ GeV$^2$. The
corresponding values for $\pi^0$- induced reactions are assumed to
be a simple average of those for $\pi^+$ and $\pi^-$- induced
reactions.}

\end{table}

\begin{table}[h]
\begin{center}
\begin
{tabular}{|c|c|c|c|c|}
  \hline



 Reaction&\multicolumn{2}{c|}{$W^2 <$ 9 GeV$^2$}&\multicolumn{2}{c|}{$W^2 >$
  9 GeV$^2$}
  \\ \cline{2-5}
&'Incident'&$<p_T^2>$&'Incident'& $<p_T^2>$\\
&momentum&(GeV$/c)^2$&momentum&(GeV$/c)^2$ \\ \hline $\pi^+
N\rightarrow \pi^+(x_F > 0) X$&&0.14&&0.18 \\ $\pi^+ N\rightarrow
\pi^-(x_F > 0) X$&1.7 GeV$/c$&0.11&2.9 GeV$/c$&0.12 \\ $\pi^+
N\rightarrow \pi^+(x_F < 0) X$&&0.28&&0.30 \\ $\pi^+ N\rightarrow
\pi^-(x_F < 0) X$&&0.15&&0.17 \\ \hline  $\pi^- N\rightarrow
\pi^-(x_F > 0) X$&&0.12&&0.15 \\$\pi^- N\rightarrow \pi^+(x_F > 0)
X$&1.3 GeV$/c$&0.10&2.4 GeV$/c$&0.11 \\ $\pi^- N\rightarrow
\pi^-(x_F < 0) X$&&0.22&&0.24 \\$\pi^- N\rightarrow \pi^+(x_F < 0)
X$&&0.12&&0.14 \\ \hline$\pi^+ N\rightarrow p X$&1.7
GeV$/c$&0.30&2.9 GeV$/c$&0.44 \\$\pi^- N\rightarrow p X$ &1.3
GeV$/c$ &0.22&2.4 GeV$/c$&0.37 \\ \hline $\pi^+ p \rightarrow
\pi^+ p$&1.7 GeV$/c$&0.47&2.9 GeV$/c$&0.85 \\  $\pi^- p
\rightarrow  \pi^- p$&1.3 GeV$/c$&0.33&2.4 GeV$/c$&0.71 \\ $\pi^+
p\rightarrow  \pi^+ p$&0.71 GeV$/c$&0.09&0.85 GeV$/c$&0.10
\\  $\pi^- p\rightarrow \pi^- p$&0.54 GeV$/c$&0.05&0.68 GeV$/c$&0.08 \\ \hline
\end{tabular}

\end{center}
\caption{The values of $<p_T^2>$ (with respect to the weak current
axis) acquired by secondary  $\pi^{\pm}$ mesons with $x_F > 0$ and
$x_F < 0$ and non-identified protons in intranuclear inelastic
interactions, and by non-identified protons in $\pi p$ elastic
scattering, at $W^2 < 9$ GeV$^2$ and $W^2 > 9$ GeV$^2$. The
corresponding values for $\pi^0$- induced reactions are assumed to
be a simple average of those for $\pi^+$ and $\pi^-$- induced
reactions.}
\end{table}


\begin{figure}
\resizebox{0.8\textwidth}{!}{\includegraphics*[bb =5 160 450
410]{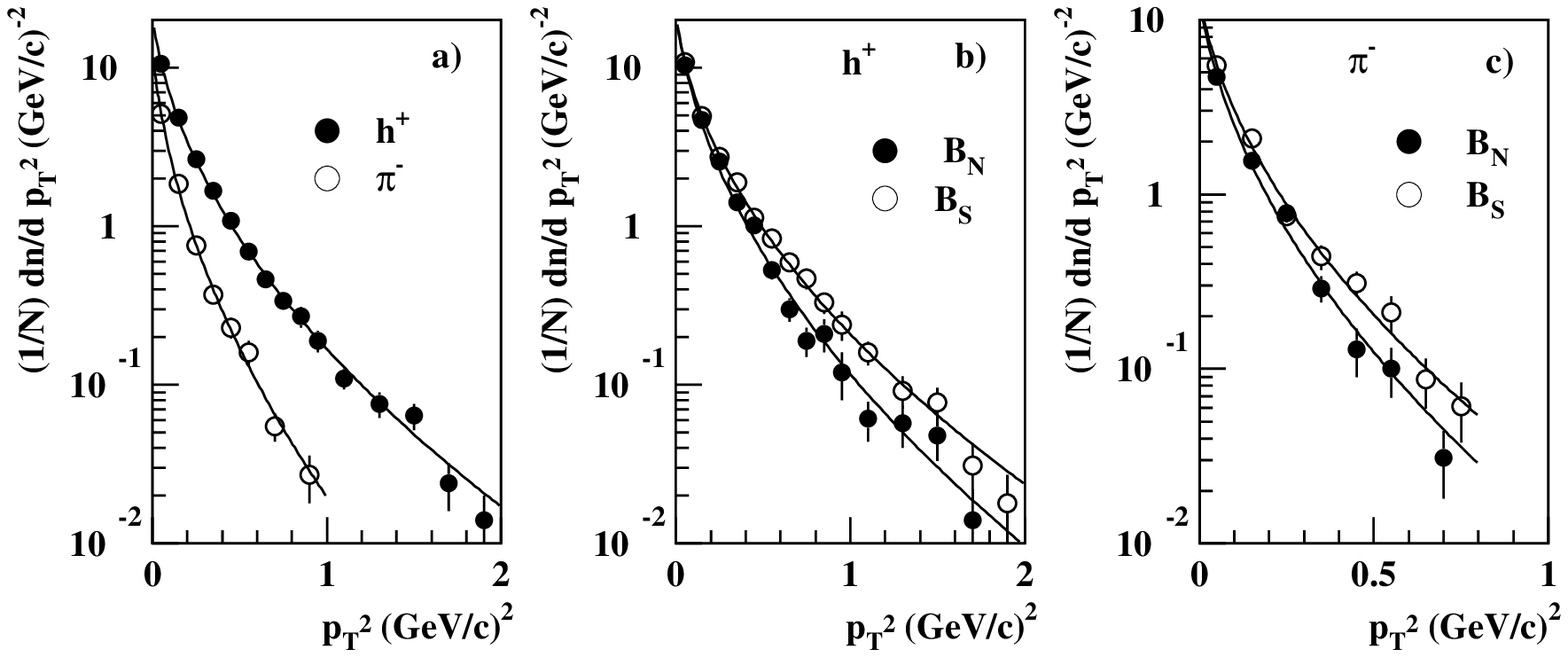}} \caption{The $p_T^2$- distributions of a) $h^+$
and ${\pi^-}$ for the whole event sample, b) $h^+$ for $B_N$ and
$B_S$ subsamples, c) ${\pi^-}$ for $B_N$ and $B_S$ subsamples. 
Lines are results of the fit (see text).}


\vspace{0.8cm}

\resizebox{0.7 \textwidth}{!}{\includegraphics*[bb=5 120 400 450]
{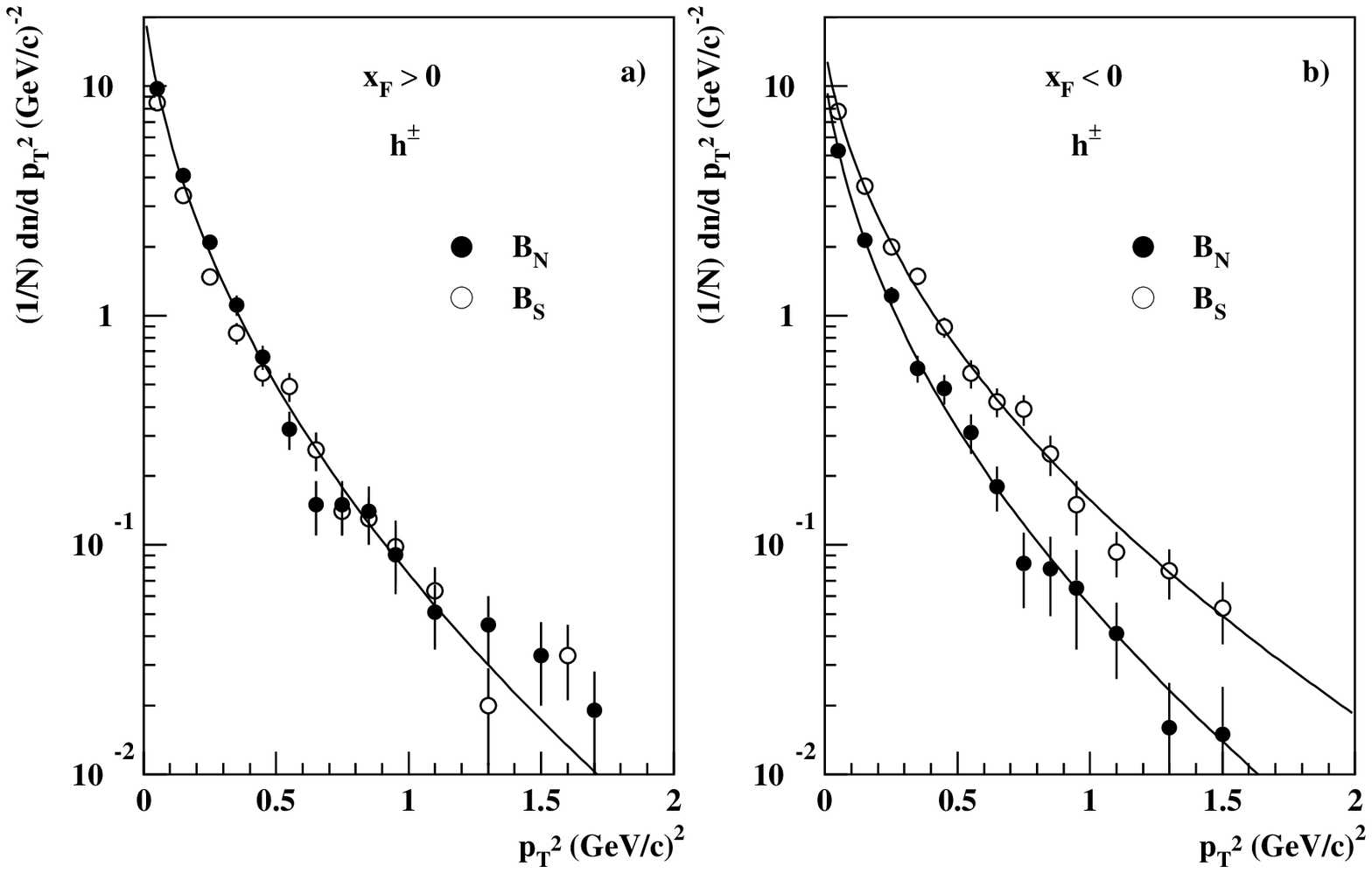}} \caption{The $p_T^2$- distributions of charged hadrons
for $B_N$ and $B_S$ subsamples at  a) $x_F > 0$ and b) $x_F < 0 $. 
Lines are results of the fit (see text).}

\end{figure}

\newpage

\begin{figure}[h]
\resizebox{.8 \textwidth}{!}{\includegraphics*[bb=5 35 500
600]{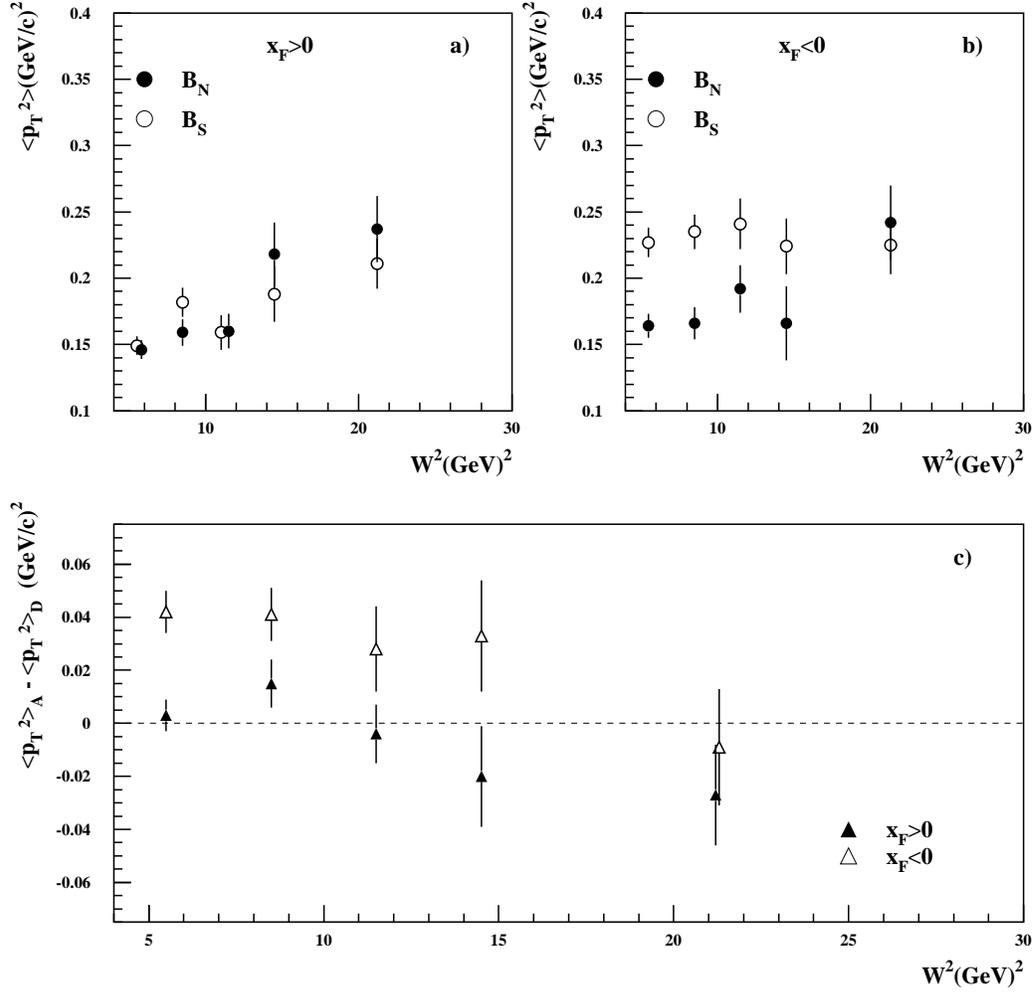}} \caption{The $W^2$- dependence of a) $<p_T^2>$ at
$x_F >$ 0 for $B_N$ and $B_S$ subsamples,  b) $<p_T^2>$ at $x_F
<$ 0 for $B_N$ and $B_S$ subsamples, c) the difference $<p_T^2>_A
- <p_T^2>_D$.}

\end{figure}


\begin{figure}[h]
\resizebox{.8\textwidth}{!}{\includegraphics*[bb=5 35 450
530]{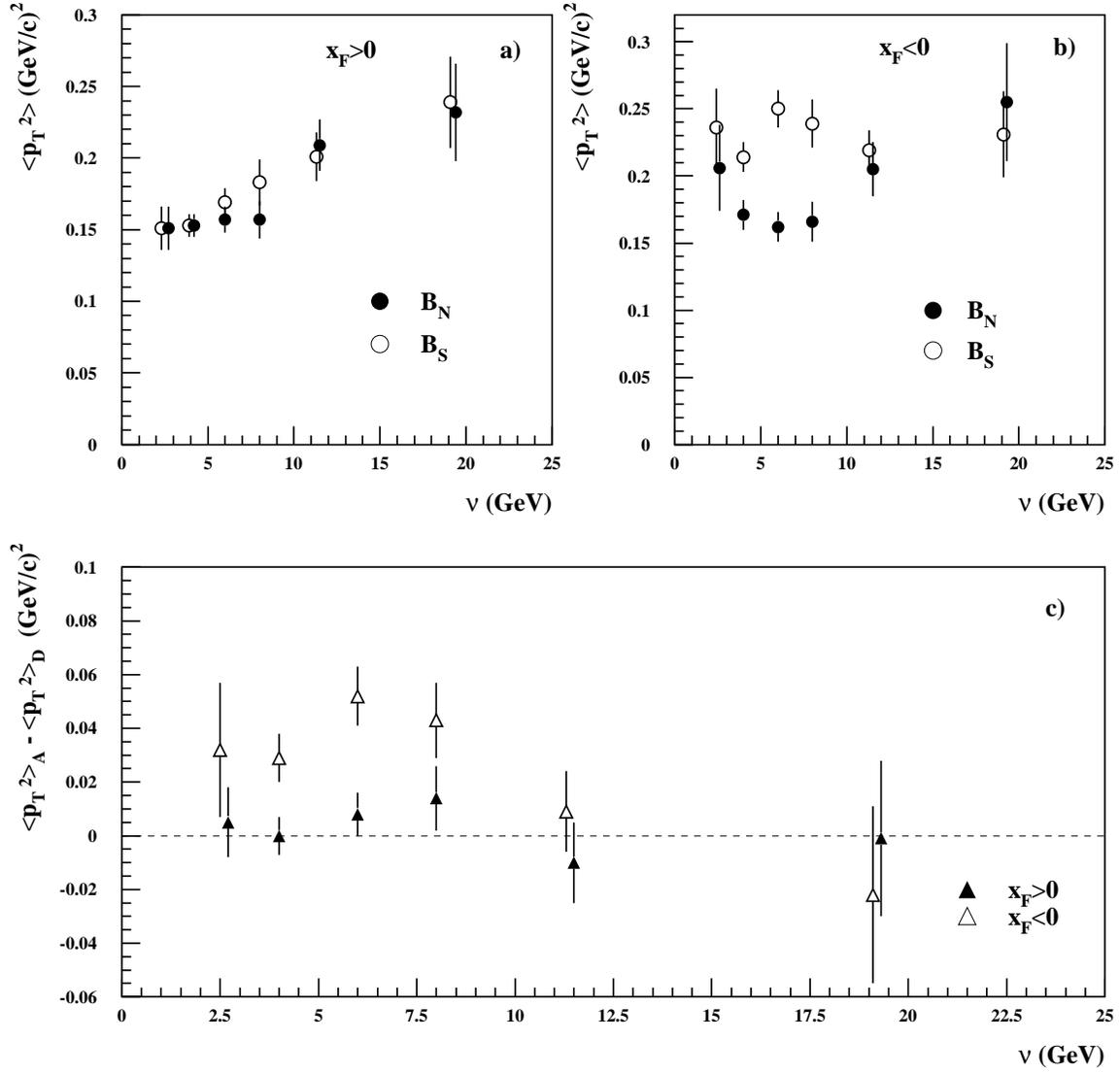}} \vspace{0.3cm} \caption{The $\nu$- dependence of
a) $<p_T^2>$ at $x_F > 0$ for $B_N$ and $B_S$ subsamples,  b)
$<p_T^2>$ at $x_F < 0$ for $B_N$ and $B_S$ subsamples, c) the
difference $<p_T^2>_A - <p_T^2>_D$.}
\end{figure}

\newpage

\begin{figure}
\resizebox{.8\textwidth}{!}{\includegraphics*[bb=5 100 450
480]{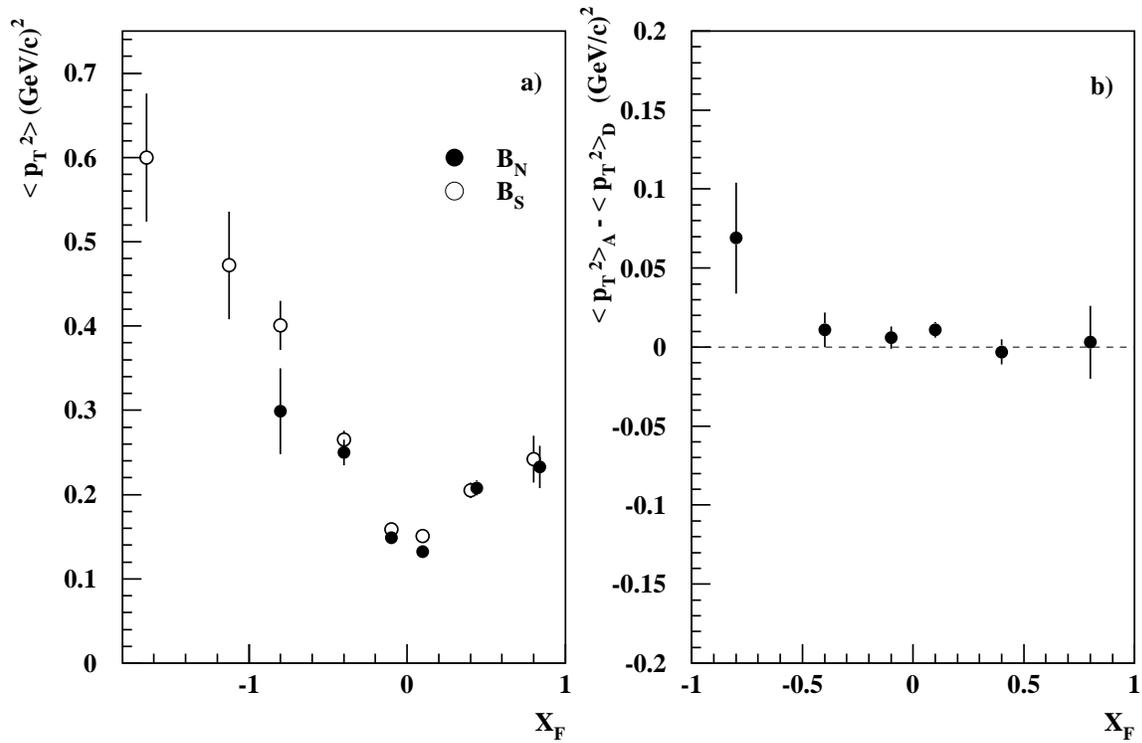}} \caption{The $x_F$- dependence of a) $<p_T^2>$ for
$B_N$ and $B_S$ subsamples,  b) the difference $<p_T^2>_A -
<p_T^2>_D$.}
\end{figure}

\begin{figure}
\resizebox{0.8\textwidth}{!}{\includegraphics*[bb=5 35 450
530]{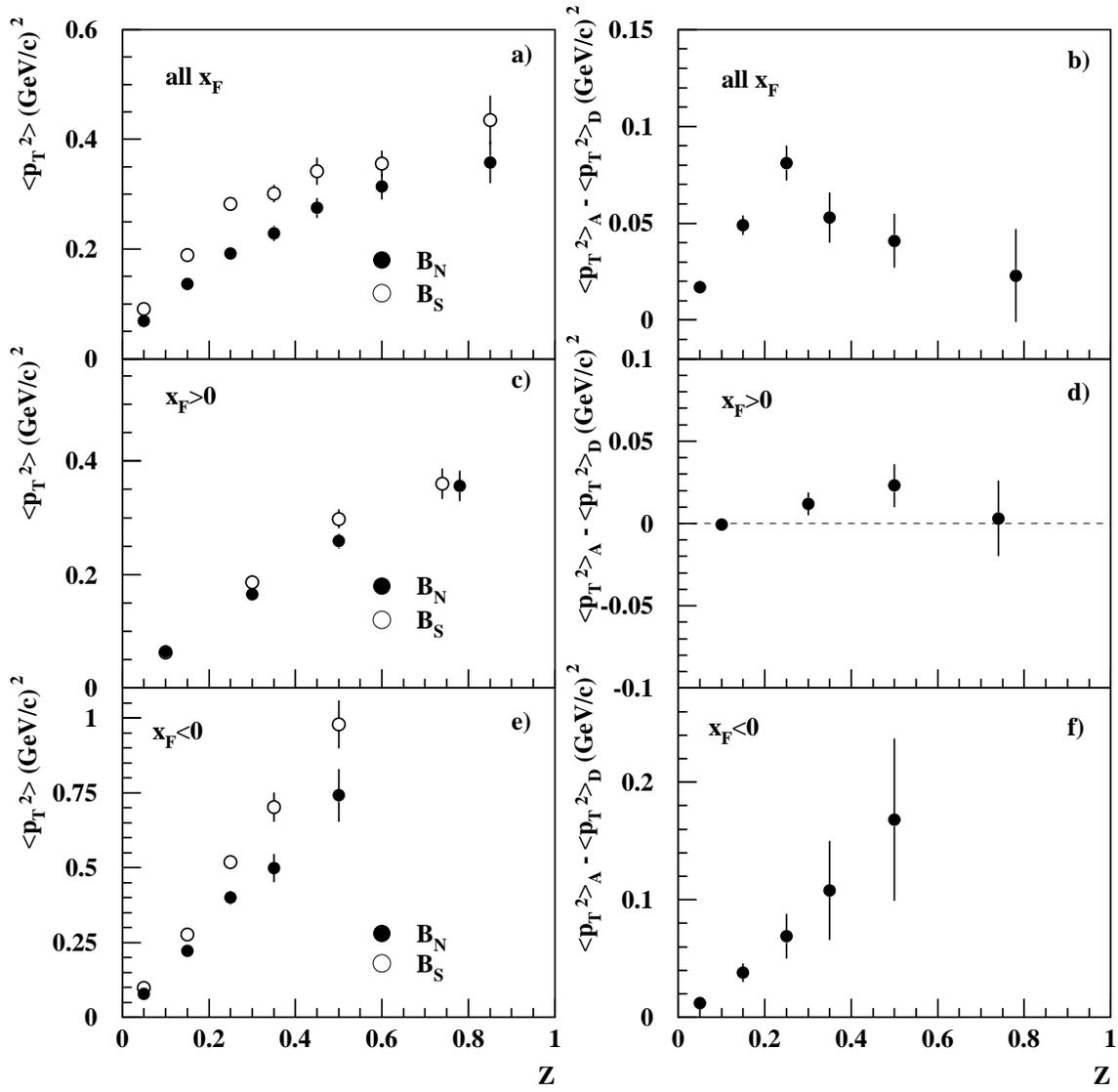}} \vspace{0.3cm} \caption{The $z$- dependence of
$<p_T^2>$ for $B_N$ and $B_S$ subsamples (a, c, e)  and of the
difference $<p_T^2>_A - <p_T^2>_D$ (b, d, f).}
\end{figure}

\newpage

\begin{figure}
\resizebox{.8\textwidth}{!}{\includegraphics*[bb=5 45 450
560]{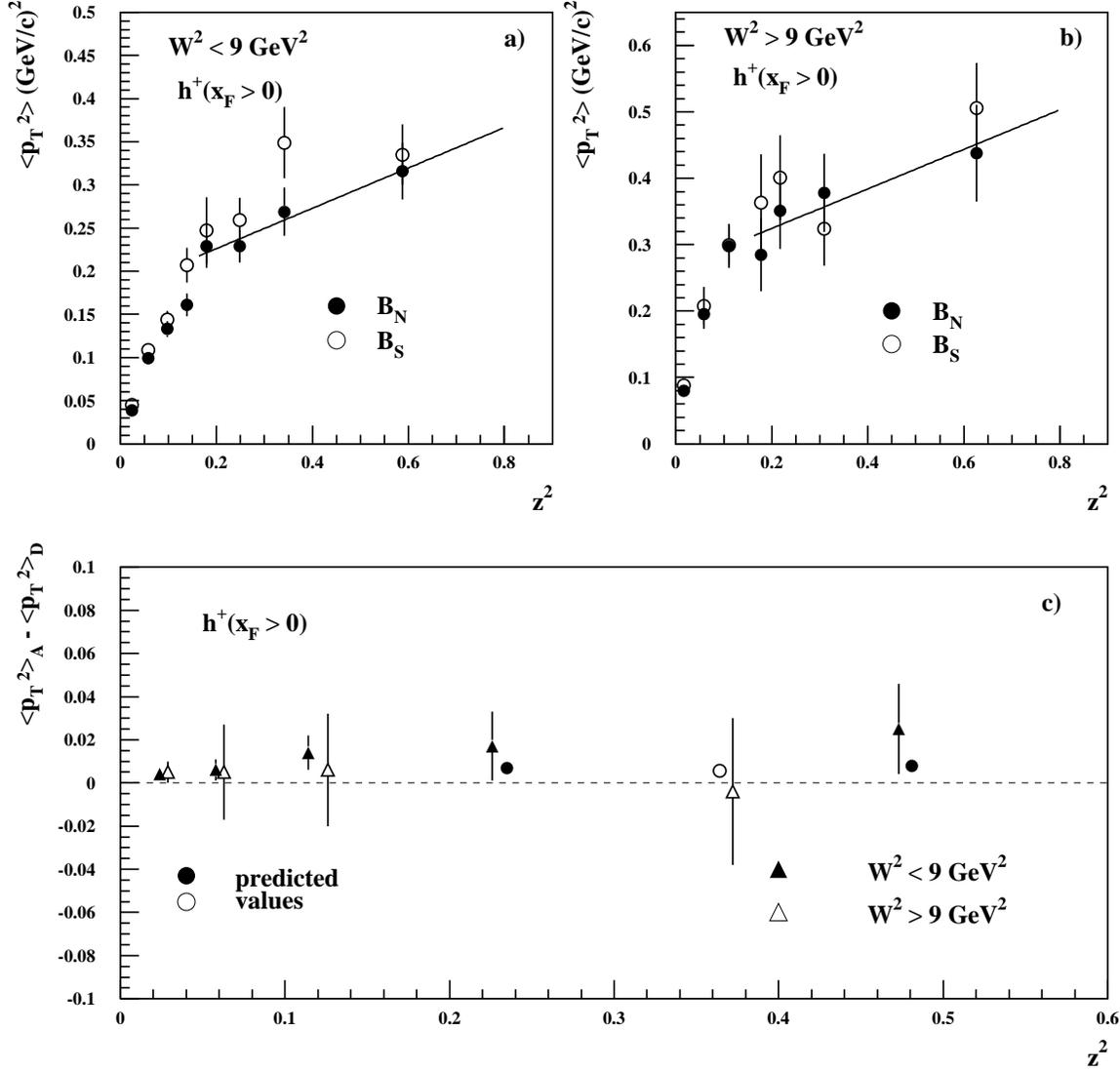}}\vspace{0.2cm} \caption{The $z^2$- dependence of:
$<p_T^2>$ of positively charged hadrons with $x_F > 0$ \, 
a) at $W^2 < 9$ GeV$^2$ and \, b) $W^2 > 9$ GeV$^2$; \, c) the
difference $<p_T^2>_A - <p_T^2>_D$. 
The lines in Figs.~7a and 7b
are the fit results (see text).}
\end{figure}

\end{document}